\documentclass[12pt,epsfig,a4]{article} 
\newcommand{\vs}{\vspace{-0.25cm}}
\usepackage{amsmath,graphicx}
\begin{document} 
\begin{center}
{\Large{\bf Three-pion exchange nucleon-nucleon potentials\\ with virtual
$\Delta$-isobar excitation}\footnote{This work 
has been supported in part by DFG and NSFC (CRC110).}  }  

\medskip

 N. Kaiser \\
\medskip
{\small Physik-Department T39, Technische Universit\"{a}t M\"{u}nchen,
   D-85747 Garching, Germany}
\end{center}
\medskip
\begin{abstract}
The nucleon-nucleon interaction arising from the exchange of three pions 
and the excitation of $\Delta(1232)$-isobars in intermediate states is 
studied. Approximating the $\Delta$-propagator by the inverse $\Delta$N 
mass-splitting, analytical expressions are derived for the spectral-functions 
of the isoscalar and isovector central, spin-spin and tensor 
NN-potentials in momentum-space. A trans- lation of the spectral-functions 
into coordinate-space potentials reveals that the main effect of these 
specific exchange and excitation mechanisms is a repulsive isoscalar central 
NN-potential.  
\end{abstract}

\section{Introduction and summary}
The interaction between two nucleons based on chiral effective field theory 
has been studied in great detail over the past two decades, for reviews see 
\cite{evgeni,evhwulf,entem,evgulf}. Most calculations employ the effective 
chiral Lagrangian formulated in terms of pions and nucleons that are chirally 
coupled with each other and to external sources. The excitation of baryon and
meson resonances is encoded in the low-energy constants of the chiral 
pion-nucleon interaction terms of higher order. Such a framework provides an 
accurate description of the empirical nucleon-nucleon phase shifts, if extended
to sufficiently high order. At the present time, calculations have been carried 
out up to order N$^4$LO in the chiral (low-momentum) expansion \cite{n4lo,bochum} 
and the incorporation of dominant contributions at order N$^5$LO is underway. 
Still, it can be argued the explicit inclusion the $\Delta(1232)$-isobar, 
the most prominent resonance in nuclear physics, allows one to resum a 
certain class of important contributions and therefore may lead to an improved 
convergence. In such a phenomenological extension of baryon chiral 
perturbation theory the delta-nucleon mass splitting, $\Delta =293\,$MeV, is 
counted as an additional small scale parameter, comparable to the typical 
momentum $p$ or the pion mass $m_\pi$. The finite-range parts of the 
$2\pi$-exchange NN-interaction as generated by the pertinent triangle and 
box diagrams with intermediate $\Delta(1232)$-isobars have been calculated 
first in ref.~\cite{2pidelta} using the Feynman diagram technique. Furthermore, 
the Bochum-Bonn group \cite{nforcedelta,isobrdelta} has worked out the 
subleading isospin-conserving corrections as well as the isospin-breaking 
corrections provided by the chirally interacting $\pi N\Delta$-system.

The purpose of the present paper is to analyze the two-nucleon interaction 
which arises from the exchange of three pions in combination with 
excitations of $\Delta(1232)$-isobars in intermediate states. In this sense 
it represents an extension of earlier works on the $3\pi$-exchange 
NN-interaction in refs.~\cite{3piexchange,subleading} to chiral effective 
field theory with explicit $\Delta(1232)$-isobar degrees of freedom.

The present paper is organized as follows. In section~2, the techniques to 
the calculate spectral-functions (or imaginary parts) of two-loop 
$3\pi$-exchange diagrams in the heavy nucleon mass limit are prepared. In 
order to facilitate an analytical treatment (of at least the angular part) of 
the involved $3\pi$-phase space integrals, the non-relativistic 
$\Delta$-propagator will be approximated by the inverse mass-splitting 
$\Delta^{-1}$. A consequence of this approximation is that the pion-induced 
excitation and deexcitation of virtual $\Delta(1232)$-isobars can be 
conveniently condensed into (symmetrized) $2\pi$- and $3\pi$-contact vertices 
with a nucleon. In section~3, analytical expressions are derived for the 
spectral-functions corresponding to the isoscalar and isovector central, 
spin-spin and tensor NN-potentials in momentum-space. This is separately done 
for five different classes of $3\pi$-exchange diagrams which are grouped 
together according to the number of intermediate $\Delta(1232)$-isobar 
excitations. For each class the resulting NN-potentials in coordinate-space 
are displayed in the distance region $1\,$fm$\,<r<2\,$fm. As a summary one finds 
that the main effect of these specific exchange and excitation mechanisms 
is a repulsive isoscalar central NN-potential $\widetilde V_C(r)$. Moreover, 
the contributions to the isovector tensor potential $\widetilde W_T(r)$ tend to 
cancel each other, while other components come out very small anyway. This 
overall result is remarkable in view of the fact that previous calculations 
of the chiral three-pion exchange NN-interaction in refs.~\cite{3piexchange,
subleading} have lead to a vanishing isoscalar central potential.

\section{Preparation}
Let us start with recalling the form of the (static) nucleon-nucleon potential
in momentum-space:
\begin{eqnarray} T_{N\!N} &=& V_C(q) + V_S(q)\,\vec \sigma_1\!\cdot\! \vec 
\sigma_2 + V_T(q) \,\vec \sigma_1\!\cdot\!\vec q \,\,\vec \sigma_2\!\cdot\!\vec 
q \nonumber \\ && + \big\{ W_C(q) + W_S(q)\,\vec \sigma_1\!\cdot\!\vec \sigma_2
+ W_T(q) \,\vec \sigma_1\!\cdot \!\vec q \,\,  \vec \sigma_2\! \cdot \!\vec q 
\, \big\}  \,\vec \tau_1\! \cdot\! \vec \tau_2\,, \end{eqnarray} 
where $q = |\vec q\,|$ denotes the momentum transfer between the initial- and
final-state nucleons. The subscripts $C, S$ and $T$ refer to the central, 
spin-spin and tensor-type components, each of which occurs in an isoscalar 
$(V_{C,S,T})$ and an isovector version $(W_{C,S,T})$. The sign-convention for 
$T_{N\!N}$ is chosen such that the usual one-pion exchange gives: $W_T(q)^{(1\pi)}
= (g_A/2f_\pi)^2(m_\pi^2+q^2)^{-1}$. The occurring physical parameters are: 
the nucleon axial-vector coupling constant $g_A=1.29$, the pion 
decay constant $f_\pi= 92.4\,$MeV, and the average pion mass $m_\pi = 138\,$MeV.

We are interested in the finite-range parts of $T_{N\!N}$ that are generated 
by certain  $3\pi$-exchange diagrams. For this purpose it is sufficient to 
calculate the imaginary parts Im$V_{C,S,T}(i\mu)$ and Im$W_{C,S,T}(i\mu)$, 
obtained by analytical continuation to time-like momentum transfer $q = i\mu$, 
with $\mu> 3m_\pi$. These imaginary parts serve also as the mass spectra 
entering a representation of the (local) coordinate-space potentials in the 
form of a continuous superposition of Yukawa functions:    
\begin{eqnarray}  \widetilde V_C(r) &=& 
-{1\over 2\pi^2 r} \int_{3m_\pi}^\Lambda\! d\mu \,\mu \,e^{-\mu r}
\, {\rm Im} V_C(i\mu)\,,  \\ \widetilde V_S(r) &=& {1\over 6\pi^2 r} 
\int_{3m_\pi}^\Lambda\! d\mu \,\mu \,e^{-\mu r} \big[\mu^2{\rm Im}V_T(i\mu) 
- 3 {\rm Im}V_S(i\mu) \big]\,, \\  \widetilde V_T(r)&=& 
{1\over 6\pi^2 r^3}\int_{3m_\pi}^\Lambda\! d\mu\,\mu\, e^{-\mu r}(3+
3\mu r+ \mu^2r^2) {\rm Im} V_T(i\mu)\end{eqnarray} 
with $\Lambda$ the spectral cutoff, set to a maximal value of $\Lambda=1.5\,$GeV 
in ref.~\cite{n4lo}. For the isovector potentials $\widetilde W_{C,S,T}(r)$ a 
completely analogous representation holds. Note that the tensor potentials 
$\widetilde V_T(r)$ and $\widetilde W_T(r)$ are accompanied by the usual tensor 
operator $3\vec \sigma_1\!\cdot \!\hat r\,\vec \sigma_2\!\cdot\!\hat r
-\vec \sigma_1\!\cdot\vec \sigma_2$.

Making use of the Cutkosky cutting rules, the imaginary parts entering 
eqs.(2,3,4) are calculated from the pertinent two-loop $3\pi$-exchange 
diagrams as integrals of the $\bar NN\!\to\! 3\pi\!\to\! \bar NN$ transition 
amplitudes over the Lorentz-invariant three-pion phase-space. In the 
center-of-mass frame this four-dimensional phase-space integral includes an 
angular part of the form:
\begin{equation} \int\!\!\!\!\int\limits_{\!\!\!\!\!\!E}\!{dx d y \over \sqrt{ 
1-x^2-y^2-z^2+2x y z}}\dots \,, \end{equation} 
with $x=\vec v\!\cdot\!\hat k_1$ and $y=\vec v\!\cdot\!\hat k_2$ the 
directional cosines of two pion-momenta $\vec k_{1,2}$. The unit-vector $\vec v$
is introduced by the four-velocity $v^\alpha = (0, i \vec v\,)$ of the heavy 
nucleon in the considered $t$-channel kinematics $\bar NN\!\!\to 3\pi\!\to\! 
\bar NN$ \cite{3piexchange}. The integration region $E$ in eq.(5) is an ellipse 
$x^2+y^2-2x yz<1-z^2$ with semi-axes $\sqrt{1+z}$ and $\sqrt{1-z}$, where  
$z=\hat k_1\!\cdot\!\hat k_2$ (see also eq.(28)). In order to allow for an 
analytical treatment of the angular integral in eq.(5) for all two-loop 
$3\pi$-exchange diagrams, we approximate the non-relativistic delta-propagator 
by the inverse delta-nucleon mass splitting $\Delta^{-1}$. In the case of a 
simple one-loop $2\pi$-exchange triangle diagram this approximation 
leads to the inequality: 
 \begin{equation} {\sqrt{\mu^2-4m_\pi^2} \over \mu \Delta} > {2\over \mu} 
\arctan{\sqrt{\mu^2-4m_\pi^2} \over 2\Delta}\,,\end{equation}
where the complete result stands on the right hand side. A comparison of the 
associated central potentials reveals that the approximation using 
$\Delta^{-1}$ leads to an overestimation by about $10-20\%$. Although 
the kinematical situation is more complex for $3\pi$-exchange, one can expect 
that the approximation using $\Delta^{-1}$ provides an upper bound for the 
NN-potentials with a similar error margin.   

Once the energy-dependence of the delta-propagator is neglected, one can 
combine the direct and crossed $\Delta(1232)$-isobar excitation to a 
$2\pi$-contact vertex as symbolized by a filled square in the left diagram of 
Fig.\,1. The corresponding transition matrix-element reads:
\begin{equation}M_{2\pi \Delta}= {i g_A^2 \over 4f_\pi^2\Delta}\Big\{\epsilon_{abc} 
\tau_c \, \vec \sigma\!\cdot\!(\vec q_a\!\times\!\vec q_b) -4\delta_{ab}\, 
\vec q_a\!\cdot\!\vec q_b\Big\}\,, \end{equation}
where $\vec q_a$ and $\vec q_b$ denote both outgoing pion-momenta. In order to 
arrive at this form the relation $T_aT_b^\dagger = (2\delta_{ab}-i\, \epsilon_{abc}
\tau_c)/3$ for the isospin (and spin) transition operators and the coupling 
constant ratio $g_{\pi N\Delta}/g_{\pi NN} = 3/\sqrt{2}$ have been used. The strong 
$\pi NN$-coupling constant follows from the Goldberger-Treiman relation as
$g_{\pi NN} = g_AM/f_\pi= 13.1$, with $M=939\,$MeV the nucleon mass. The 
delta-nucleon mass splitting $\Delta$ has the well-known value 
$\Delta = 293\,$MeV. 

Fig.\,1 shows on the right also a diagram with a $3\pi$-contact vertex which 
arises from an additional direct coupling of the pion to the 
$\Delta(1232)$-isobar. Summing over the six permutations of the isospin-indices 
$(a,b,c)$ the corresponding transition matrix-element reads:
\begin{eqnarray}M_{3\pi \Delta\Delta}&=& {g_A^3 \over 40 f_\pi^3 \Delta^2} \Big\{ 
-75 \epsilon_{abc} \,\vec q_a\!\cdot\!(\vec q_b\!\times\!\vec q_c) + \vec q_a\!
\cdot\! \vec q_b\,\vec \sigma\!\cdot \!\vec q_c\, (18 \delta_{ab}\tau_c-7  
\delta_{ac}\tau_b-7\delta_{bc}\tau_a)\nonumber \\ && + \vec q_a\!\cdot\! \vec 
q_c\,  \vec \sigma\!\cdot\! \vec q_b\, (18 \delta_{ac}\tau_b-7  \delta_{ab}\tau_c
-7\delta_{bc}\tau_a)+ \vec q_b\!\cdot\!\vec q_c\,  \vec \sigma\!\cdot\! \vec 
q_a\, (18 \delta_{bc}\tau_a-7 \delta_{ac}\tau_b-7\delta_{ab}\tau_c)\Big\}\,, 
\end{eqnarray}
where $\vec q_a, \vec q_b, \vec q_c$ denote outgoing pion-momenta. In order to
fix the $\Delta\Delta\pi$-vertex we use the coupling constant ratio 
$g_{\pi \Delta \Delta}/g_{\pi NN}=1/5$ of the quark-model. Furthermore, the relation 
for the isospin (transition) operators, $T_a \Theta_b T_c^\dagger = (5i\, 
\epsilon_{abc}-\delta_{ab}\tau_c +4 \delta_{ac}\tau_b -\delta_{bc}\tau_a)/3$, with 
$\Theta_b$ the $4\times 4$ $\Delta$-isospin matrices, has been employed 
together with an analogous relation for the spin (transition) operators.
  
\begin{figure}
\begin{center}
\includegraphics[scale=0.55,clip]{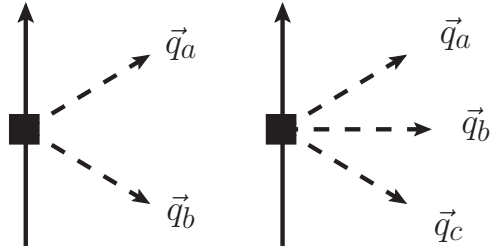}
\end{center}
\vspace{-.5cm}
\caption{Contact-vertices of two and three pions with a nucleon arising from 
the excitation of intermediate $\Delta(1232)$-isobars.}
\end{figure}

While the decomposition into isoscalar and isovector contributions is obvious 
from the isospin factor $V +W \, \vec \tau_1\!\cdot\! \vec \tau_2$ of a 
$3\pi$-exchange diagram, a certain technique is needed in order to separate 
the \\ spin-spin and tensor-like components. First, one  removes from a 
spin-dependent expression of the form $\vec \sigma_1 \!\cdot\! \vec A \, \vec 
\sigma_2\! \cdot\!\vec B$ the Pauli-matrices $\vec\sigma_1$ and $\vec\sigma_2$. 
The spin-spin part Im$V_S$ is then obtained by con- tracting the Lorentz-tensor 
$A_\alpha B_\beta $ with the projector ${1\over 2}(v^\alpha v^\beta +\mu^{-2} 
q^\alpha q^\beta - g^{\alpha \beta})$, and the combination Im$(\mu^2V_T-V_S)$ 
including the tensor part follows finally by contracting $A_\alpha B_\beta $ 
with the projector $\mu^{-2} q^\alpha q^\beta$. After these manipulations the
relevant expressions are given in terms of Lorentz-scalars, which can be easily 
translated from the $s$-channel ($NN\!\to\! NN$) into the $t$-channel 
($\bar NN\!\to\! 3\pi\!\to\! \bar NN$). 
\section{Calculation of spectral functions and r-space potentials}
In this section, we present the analytical expressions for the 
spectral-functions pertaining to the isoscalar and isovector central, 
spin-spin and tensor NN-potentials in momentum-space. This is done separately 
in five subsections for the different classes of $3\pi$-exchange diagrams, 
ordered according to the number of intermediate $\Delta(1232)$-isobar 
excitations. In all cases the technical details are omitted and only the final 
results for the non-vanishing contributions are given. The calculated 
spectral-functions are then used to construct the corresponding NN-potentials 
in coordinate-space. These are presented in five figures displaying their 
dependence on the distance $r$ in the region $1\,$fm$\,<r<2\,$fm. 

\subsection{Single $\Delta$-excitation of one nucleon}
We start with the $3\pi$-exchange diagrams with a single 
$\Delta(1232)$-excitation of one nucleon. The corresponding diagrams are 
compiled in ref.~\cite{subleading}, where they have been grouped into the classes
X, XI, XII, XIII and XIV. The subleading $2\pi$-contact vertex treated in 
ref.~\cite{subleading} includes $M_{2\pi \Delta}$ as a special case, namely by 
setting the low-energy constants to the values: $c_1=0,\, c_2=-c_3 = 2c_4=
g_A^2/2\Delta$. Therefore, it suffices to evaluate the $dw$-integrals in 
eqs.(7-22) of ref.~\cite{subleading} for these parameter values. It is 
convenient to introduce the dimensionless variable $u = \mu/m_\pi>3$ and the 
auxiliary functions:
\begin{equation} R(u)= (u-1)\sqrt{(u-3)(u+1)}\,, \qquad 
L(u)  = \ln{\sqrt{u-3}+\sqrt{u+1}\over 2}\,. \end{equation}
The contributions proportional to $g_A^4$ which arise from the classes X, XI 
and XII read:
\begin{equation} {\rm Im}V_S = {g_A^4 m_\pi^5\over (4f_\pi)^6 \pi^2 \Delta} 
\bigg\{{R(u)\over 48}\bigg[100-{27\over u^3}- {50\over u^2}-{151\over u}
+185u-14u^2-7u^3\bigg] + 2L(u) \bigg[{2\over u^3} + {10\over u} - 9 u\bigg]
\bigg\}\,, \end{equation}
\begin{equation} {\rm Im}(\mu^2V_T-V_S)={g_A^4m_\pi^5\over (4f_\pi)^6 
\pi^2 \Delta} \bigg\{{R(u)\over 24}\bigg[u^3+2u^2-39u-12+{65\over u}-{50\over 
u^2}-{27\over u^3} \bigg]+ 4L(u) \bigg[{2\over u^3} - {10\over u} +3 u\bigg]
\bigg\}\,, \end{equation}
\begin{equation} {\rm Im}W_S = {g_A^4 m_\pi^5\over (4f_\pi)^6 \pi^2 \Delta} 
\bigg\{{R(u)\over 72}\bigg[{135\over u^3}+{58\over u^2}-{277\over u}-36 
+147u -10u^2-5u^3\bigg] + 4L(u) \bigg[{2\over 3u^3} + {2\over 3u}-u\bigg]
\bigg\}\,, \end{equation}
\begin{eqnarray} {\rm Im}(\mu^2W_T-W_S)&=&{g_A^4m_\pi^5\over (4f_\pi)^6 
\pi^2 \Delta} \bigg\{R(u)\bigg[{15\over 4u^3}+{29\over 18u^2}+{77\over 9u}
-{13\over 2}-{u\over 4}+{2u^2\over 9}+{u^3\over 9}\nonumber \\ &&
-{1\over 2(u+1)}-{29\over 6(u-1)} \bigg]+ 8L(u) \bigg[{2\over 3u^3}
+{11\over 3u}+{u\over u^2-1}\bigg]\bigg\}\,. \end{eqnarray}
The additional contributions proportional to $g_A^6$ from the 
classes XIII and XIV take the form:
\begin{equation} {\rm Im}W_C = {g_A^6 m_\pi^5\over (4f_\pi)^6 \pi^2 \Delta} 
\bigg\{{2R(u)\over 3}\big[u-2u^2-u^3-4\big]+ 16L(u) \bigg[{1\over u}-4u+u^3
\bigg]\bigg\}\,, \end{equation}
\begin{equation} {\rm Im}V_S = {g_A^6 m_\pi^5\over (4f_\pi)^6 \pi^2 \Delta} 
\bigg\{{R(u)\over 48}\bigg[13 u^3+26 u^2-371 u-76+{493\over u}-{58\over u^2}
-{135\over u^3}\bigg] + 2L(u) \bigg[15u-{22\over u} - {2\over u^3}\bigg]
\bigg\}\,, \end{equation}
\begin{equation} {\rm Im}(\mu^2V_T-V_S)={g_A^6m_\pi^5\over (4f_\pi)^6 
\pi^2 \Delta} \bigg\{{R(u)\over 24}\bigg[5u^3+10u^2-3u-252-{443\over u}
-{58\over u^2}-{135\over u^3} \bigg]+ 4L(u) \bigg[3u+{22\over u} - {2\over u^3}
\bigg]\bigg\}\,, \end{equation}
\begin{eqnarray} {\rm Im}W_S &=& {g_A^6 m_\pi^5\over (4f_\pi)^6 \pi^2 \Delta} 
\bigg\{{R(u)\over 144}\bigg[284-{189\over u^3}-{158\over u^2}-{1249\over u} 
+1543u -210u^2-105u^3\bigg] \nonumber \\ && +2 L(u) \bigg[{2\over 3u^3} 
+ {12\over u}-17u+2u^3\bigg]\bigg\}\,, \end{eqnarray}
\begin{eqnarray} {\rm Im}(\mu^2W_T-W_S) &=& {g_A^6 m_\pi^5\over (4f_\pi)^6 \pi^2 
\Delta} \bigg\{{R(u)\over 72}\bigg[972-{189\over u^3}-{158\over u^2}+{119
\over u} +1503u -178u^2-89u^3\bigg] \nonumber \\ && +4 L(u) \bigg[{2\over 3u^3} 
- {22\over 3u}-13u\bigg]\bigg\}\,. \end{eqnarray}
As indicated by the prefactor $m_\pi^5/\Delta$ all the terms in eqs.(10-18) are 
of fourth power in small momenta. Thus they are counted of order N$^3$LO in 
the chiral effective field theory with explicit $\Delta(1232)$-isobars. Note 
also that the isoscalar central component Im$V_C$ vanishes identically due to 
the exact cancellation of the contributions from class XIII and XIV
(see eq.(18) in ref.~\cite{subleading}).  
  
The resulting local NN-potentials in coordinate-space are displayed by Fig.\,2 
in the distance region   $1\,$fm$\,<r<2\,$fm. In order to obtain curves with 
less rapid decrease in $r$, we have divided out a Yukawa function 
$\exp(-3m_\pi r)/r$ with the decay-length $(3m_\pi)^{-1} = 0.48\,$fm. Such an 
unscaled Yukawa potential has a strength of $(24.1, 5.65, 1.49)$\,MeV at 
distances $r=(1.0, 1.5, 2.0)$\,fm, respectively. One observes from Fig.\,2 
that the strongest component is an attractive isovector tensor potential 
$\widetilde W_T(r)$, followed by a repulsive isovector central potential 
$\widetilde W_C(r)$ of about half that magnitude. The spin-spin potentials, 
$\widetilde V_S(r)$ and $\widetilde W_S(r)$, turn out to be  particularly weak. 
\begin{figure}
\begin{center}
\includegraphics[scale=0.5,clip]{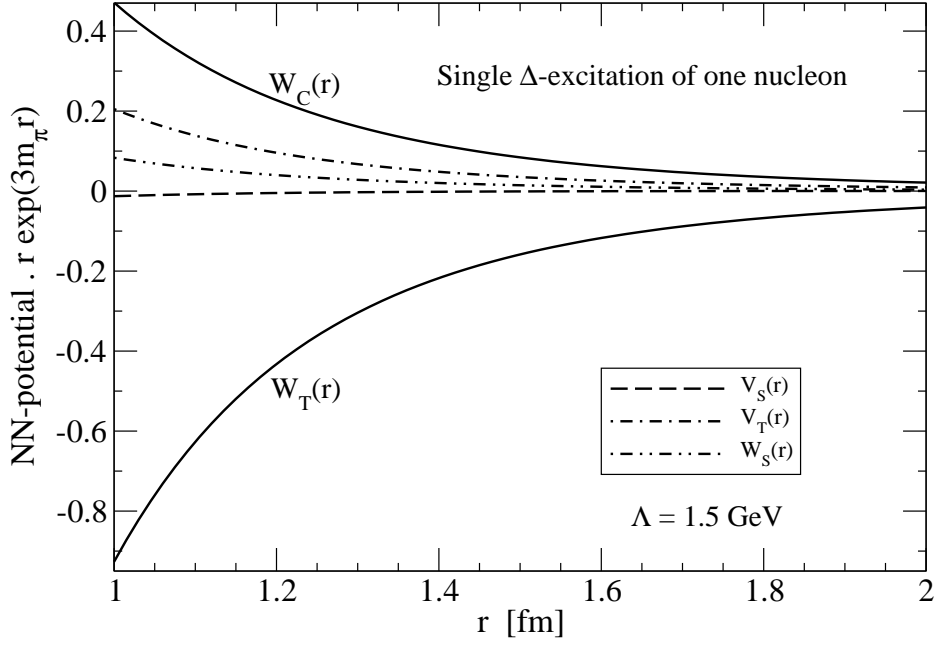}
\end{center}
\vspace{-.5cm}
\caption{NN-potentials arising from $3\pi$-exchange with single 
$\Delta$-excitation of one nucleon.}
\end{figure}
\subsection{Single $\Delta$-excitation of both nucleons}
\begin{figure}
\begin{center}
\includegraphics[scale=0.5,clip]{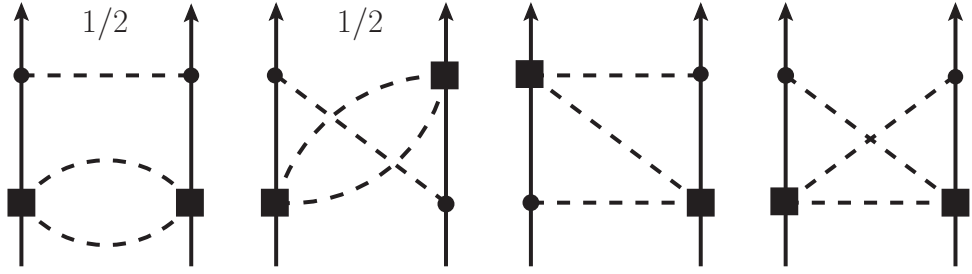}
\end{center}
\vspace{-.5cm}
\caption{Diagrams related to $3\pi$-exchange with single 
$\Delta$-excitation of each nucleon. Upside-down reflected diagrams 
are not shown. Appropriate symmetry factors $1/2$ are indicated.}
\end{figure}

Next, we come to the $3\pi$-exchange diagrams with a single 
$\Delta(1232)$-excitation of both nucleon. A representative set of two-loop 
diagrams with pion-bubbles and pions running in zigzag is shown in Fig.\,3. 
The two-nucleon irreducible contributions coming from the planar and crossed 
bubble diagrams add up to the following results for the spectral-functions: 
\begin{equation}{\rm Im}W_C = -{2g_A^6m_\pi^6\over  3\pi^3(4f_\pi)^6 \Delta^2} 
\int\limits_2^{u-1}\!dw\,(w^2-4)^{3/2}\sqrt{\lambda(w)}\,, \end{equation}
\begin{eqnarray} {\rm Im}V_S&=& {g_A^6m_\pi^6\over 12 \pi^3(4f_\pi)^6\Delta^2}
\int\limits_2^{u-1}\!dw\,{(w^2-4)^{3/2}\over u^4 \sqrt{\lambda(w)}}\Big[(u^2-1)^3
(5u^2+1) \nonumber \\ && +4w^2(1+2u^2+5u^4-2u^6)-2w^4(3+5u^2)+4(1+u^2)w^6-w^8
\Big]\,, \end{eqnarray}
\begin{equation} {\rm Im}(\mu^2V_T-V_S)={g_A^6m_\pi^6\over 12 \pi^3(4f_\pi)^6
\Delta^2} \int\limits_2^{u-1}\!dw\,(w^2-4)^{3/2}\sqrt{\lambda(w)}\,\bigg[
{2\over u^2}(7w^2+1)-{(w^2-1)^2\over u^4}-1\bigg]\,, \end{equation}
where $\lambda(w) = w^4+u^4+1-2(w^2u^2+w^2+u^2)$ denotes the (kinematical) 
K\"allen function. The dimensionless integration variable $w$ introduced here 
is the invariant mass of a pion-pair, divided by the pion mass $m_\pi$. 

The other diagrams in Fig.\,3 with pions running in zigzag lead to contributions 
to the spectral-function which can be given in closed  analytical form:
\begin{equation}{\rm Im}V_C = {g_A^6m_\pi^6\over 35\pi(4f_\pi)^6 \Delta^2}\,
(u-3)^3\bigg[{3\over u}+3-12u-9u^2-u^3\bigg]\,, \end{equation}
\begin{equation}{\rm Im}W_S = {g_A^6m_\pi^6(u-3)^2\over 70\pi(4f_\pi)^6 \Delta^2}
\bigg[2u^4+12u^3-21u^2-{83u\over 2}+80+{150\over u}-{125\over 3u^2}
-{125\over 2u^3}\bigg]\,, \end{equation}
\begin{equation}{\rm Im}(\mu^2W_T-W_S)= {g_A^6m_\pi^6(u-3)\over 70\pi(4f_\pi)^6  
\Delta^2}\bigg[{11u^5\over 3}+11u^4+2u^3-169u^2+88u+ 33 +{962\over 3u}+
{125\over u^2}+{375\over u^3}\bigg]\,, 
\end{equation}
and further contributions which can only be reduced to double-integrals of the form:
\begin{equation}{\rm Im}W_C = -{g_A^6m_\pi^6u^2\over (8\pi f_\pi^2)^3 \Delta^2}
\int\!\!\!\!\int\limits_{\!\!\!\!\!z^2<1}\!d\omega_1d\omega_2\, k_1 k_2 \sqrt{1-z^2}
\arcsin(z)\,,\end{equation}
\begin{eqnarray} {\rm Im}V_S&=& {g_A^6m_\pi^6\over (8\pi f_\pi^2)^3\Delta^2}
\int\!\!\!\!\int\limits_{\!\!\!\!\!z^2<1}\!d\omega_1d\omega_2\,\bigg\{\omega_1^2
(9\omega_2u-\omega_2^2-9u^2-1)+{3\omega_1\over 2}\big[6u+6u^3-\omega_2(1+8u^2)
\big]\nonumber \\ && -{1\over 8}(9u^4+18u^2+5)+{z k_2\over k_1}\Big[
\omega_1^3(\omega_2-4u)+\omega_1^2(2+2u^2-7\omega_2u)+2\omega_1(2u+\omega_2)
\nonumber \\ && -2-2u^2+4\omega_2 u\Big]+{3 \arcsin(z)\over k_1k_2\sqrt{1-z^2}}
\bigg[\omega_1^3u(2 \omega_2u -u^2-1)+{\omega_1^2\over 2}\Big(5\omega_2^2 u^2- 
\omega_2u (7+11u^2) \nonumber \\ &&   +1+4u^2+3u^4\Big)+{\omega_1\over 8}
\Big(\omega_2(5+16u^2+15u^4)+2u-12u^3-6u^5\Big)+{(u^4-1)(u^2+3)\over 16}
\bigg]\bigg\}\,,\nonumber \\ &&  \end{eqnarray}
\begin{eqnarray} {\rm Im}(\mu^2V_T-V_S)&=& {g_A^6m_\pi^6\over (8\pi f_\pi^2)^3
\Delta^2}\int\!\!\!\!\int\limits_{\!\!\!\!\!z^2<1}\!d\omega_1d\omega_2\,\bigg\{
2\omega_1^2(10\omega_2 u-\omega_2^2-6u^2-2)-3u^2(1+u^2) \nonumber \\ &&   
+\omega_1\big[10u+18u^3-3\omega_2(1+7u^2)\big] +{z k_2 \over k_1}\Big[
\omega_1^3(2\omega_2-7u)-u^2+\omega_2u\nonumber \\ && +\omega_1^2(3+10u^2
-13\omega_2 u)+\omega_1(2+3u^2)(2\omega_2-u)\Big]  
+{3 \arcsin(z)\over k_1k_2\sqrt{1-z^2}}\nonumber \\ && \times (u^2-2\omega_1u 
+1)(u^2-2\omega_2u+1)\bigg[{u^2\over 4}+\omega_1^2+{\omega_1\over 4}(5\omega_2 
-6u) \bigg]\bigg\}\,.\end{eqnarray}

\begin{figure}
\begin{center}
\includegraphics[scale=0.5,clip]{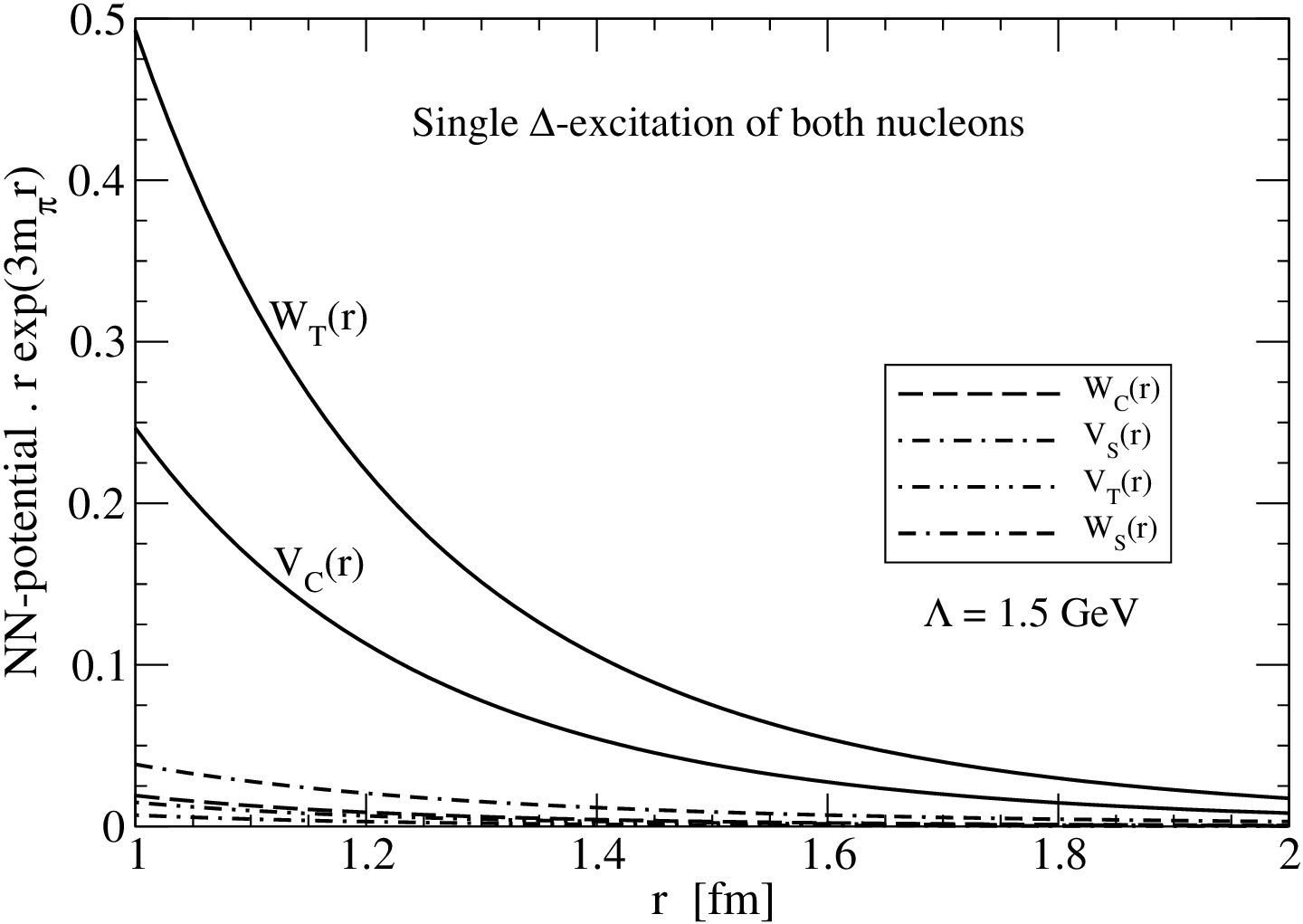}
\end{center}
\vspace{-.5cm}
\caption{NN-potentials arising from $3\pi$-exchange with single 
$\Delta$-excitation of both nucleons.}
\end{figure}

The magnitudes of the pion-momenta $\vec k_{1,2}$, divided by $m_\pi$, and their 
scalar-product are given by:
\begin{equation} k_1 = \sqrt{\omega_1^2 -1}\,, \qquad k_2=\sqrt{\omega_2^2-1}\,, 
 \qquad z\, k_1k_2 = \omega_1\omega_2-u(\omega_1+\omega_2)+ {u^2+1 \over 2} \,,
\end{equation}
where the last relation follows from energy conservation. The upper and lower 
limits of the $\omega_2$-integration are $\omega_2^\pm = {1\over 2}(u-\omega_1
\pm k_1 \sqrt{u^2-2\omega_1 u -3}/ \sqrt{u^2-2\omega_1 u +1}\,)$, with 
$\omega_1$ lying in the interval $1<\omega_1<(u^2-3)/2u$. Note that the 
non-algebraic result $\arcsin(z)/\sqrt{1-z^2}$ from the angular integral in 
eq.(5) prohibits a further step in the analytical integration over the 
$3\pi$-phase space. In the chiral limit $m_\pi \to 0$ the double-integrals in 
eqs.(25,26,27) have the values $-u^4 \pi^2/2520,\, -u^6(\pi^2/7+17/24)/360$ 
and $u^6(31/24-\pi^2/7)/720$, thus these spectral-functions become 
proportional to $\mu^6/\Delta^2$. 

Fig.\,4 shows the resulting NN-potentials in $r$-space, multiplied again with 
$r \exp(3m_\pi r)$. One observes as the largest component a repulsive isovector 
tensor potential $\widetilde W_T(r)$, and for the first time, a repulsive 
isoscalar central potential $\widetilde V_C(r)$ of about half that strength.
Note that the underlying spectral-functions, written in eqs.(22,23,24) have 
a rather simple analytical form. Other components of the NN-interaction turn 
out to be very small for this class of $3\pi$-exchange diagrams.     
\subsection{Double $\Delta$-excitation of one nucleon}

\begin{figure}
\begin{center}
\includegraphics[scale=0.5,clip]{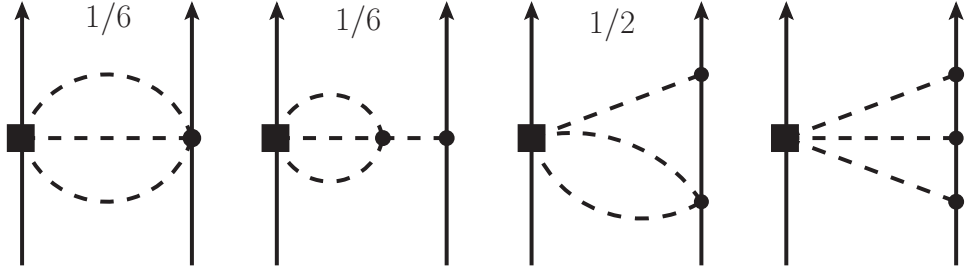}
\end{center}
\vspace{-.5cm}
\caption{Diagrams related to $3\pi$-exchange with double $\Delta$-excitation 
of one nucleon. Left-right reflected diagrams are not shown. Appropriate 
symmetry factors, $1/6$ and $1/2$, are indicated. The third diagram including a 
Weinberg-Tomozawa $2\pi$-contact vertex vanishes.}
\end{figure}

Now we consider the diagrams in Fig.\,5 including (on the left) a double 
$\Delta(1232)$-excitation of one nucleon. The first two diagrams involving the 
leading order chiral $3\pi$-contact vertex and the chiral $\pi\pi$-interaction 
give rise only to isovector spin-dependent contributions (proportional to 
$g_A^A$) of the form: 
\begin{eqnarray} {\rm Im}W_S &=& {g_A^4m_\pi^6\over 360\pi^3 (4f_\pi u)^6\Delta^2 
}\int\limits_2^{u-1}\!dw\,\sqrt{w^2-4}\,\lambda^{3/2}(w) \nonumber \\ && \times
\bigg[3+78u^2(2-w^2) +{2\over w^2}(12u^2-11u^4-1)-w^4-11u^4\bigg]\,,
\end{eqnarray}
\begin{eqnarray} {\rm Im}(\mu^2W_T-W_S)&=& {g_A^4m_\pi^6\over 72\pi^3(4 f_\pi u)^6
\Delta^2 }\int\limits_2^{u-1}\!dw\,\sqrt{(w^2-4)\lambda(w)} \bigg[
u^2-3w^2+3+{4(1-w^2)\over u^2-1}\bigg]\nonumber \\ && \times \bigg[w^4(1-43u^2)
-w^6+w^2(u^2+3)(47u^2+1)\nonumber \\ && -3u^6-79u^4-89u^2-5+{2\over w^2}(u^2-1)^2
(1-3u^2)\bigg]\,.\end{eqnarray}
Interestingly, the third diagram in Fig.\,5 involving the Weinberg-Tomozawa 
$2\pi$-contact vertex leads to vanishing spectral-functions Im$W_{S,T}=0$.
This zero result is not obvious and it is obtained after the third step of 
the analytical integration over the pion-energy $\omega_2$. Hence, one is left 
with the right diagram in Fig.\,5 involving three ordinary $\pi N$-couplings. 
The corresponding contributions are proportional to $g_A^6$ and the 
double-integral representations of the spectral-functions read: 
\begin{equation}{\rm Im}V_C = -{45g_A^6m_\pi^6u^2\over 2(8\pi f_\pi^2)^3 
\Delta^2} \int\!\!\!\!\int\limits_{\!\!\!\!\!z^2<1}\!d\omega_1d\omega_2\,
k_1 k_2 \sqrt{1-z^2} \arccos(-z)\,,\end{equation}
\begin{eqnarray} {\rm Im}W_S &=& {g_A^6m_\pi^6\over (8\pi f_\pi^2)^3\Delta^2}
 \int\!\!\!\!\int\limits_{\!\!\!\!\!z^2<1}\!d\omega_1d\omega_2\,\bigg\{
{\omega_1^3\omega_2\over 3}+{\omega_1^2\over 5}(5\omega_2 u-2\omega_2^2-6)
-{2\over 5}+2\omega_1\Big[u+\omega_2\Big({1\over 3}-u^2\Big)\Big]
\nonumber \\ && +{z k_2\over 6k_1}\bigg[ \omega_1^3(9u-4\omega_2)-
{13\omega_1^4 \over 5}+{\omega_1^2\over 5}(9+47\omega_2^2-60u^2-105\omega_2u)
+5\omega_1(2\omega_2+3u)\nonumber \\ && +15\omega_2 u-{56+47\omega_2^2\over 5}
\bigg]+z^2(\omega_2^2-1)\Big[{3\over 5}(2\omega_1^2+3)-\omega_1(\omega_2+3u)\Big]
-{k_1\over 3}(z k_2)^3 \nonumber \\ && +{\arccos(-z)\over k_1k_2\sqrt{1-z^2}} 
\bigg[\omega_1^3u(2\omega_2u-u^2-1) +\omega_1^2\Big(6u^2\omega_2^2-4\omega_2
(2u^3+3u)+u^4 \nonumber \\ && +{9u^2+5\over 2}\Big)+{\omega_1\over 4}\Big(
2\omega_2(4u^4+15u^2+5) -u^5-16u^3-7u\Big)+{u^4-5 \over 4}+u^2\bigg] \bigg\}\,,
\end{eqnarray}
\begin{eqnarray} {\rm Im}(\mu^2W_T-W_S) &=& {g_A^6m_\pi^6\over (8\pi f_\pi^2)^3
\Delta^2} \int\!\!\!\!\int\limits_{\!\!\!\!\!z^2<1}\!d\omega_1d\omega_2\,\bigg\{
{2\omega_1\over 3}(5u-\omega_1-\omega_2)(2+\omega_1\omega_2)-4u^2\omega_1
\omega_2  \nonumber \\ &&  +{z k_2\over 3k_1}\Big[5u(3u-
\omega_2)+5\omega_1(4\omega_2-u)+\omega_1^2(12-15u^2-13\omega_2u +6\omega_2^2)
 \nonumber \\ && +\omega_1^3(5u-2\omega_2)\Big] +2z^2\omega_1(\omega_2^2-1)
(\omega_1+\omega_2-5u)-{(2z k_2)^3\omega_1^2\over 3k_1}\nonumber \\ && +
{\arccos(-z)\over k_1k_2\sqrt{1-z^2}}\Big[(1+u^2)(u-2\omega_1-2\omega_2)+4
\omega_1 \omega_2 u\Big]\nonumber \\ && \times\bigg[{3u \over 4}(1+u^2)+ 
\omega_1^2 u+{\omega_1\over 2}(6\omega_2 u-5-7u^2)\bigg]\bigg\}\,.\end{eqnarray}
We note as an aside that in the chiral limit $m_\pi \to 0$ the double-integrals 
in eqs.(31,32,33) have the values $u^4 \pi^2/1260,\, u^6(\pi^2-17/48)/540$, 
and $u^6(\pi^2/7+31/480)/108$. Hence, these spectral-functions become again 
proportional to $\mu^6/\Delta^2$. It should be pointed out that the 
spin-isospin structure of the $3\pi$-contact vertex $M_{3\pi\Delta\Delta}$ in 
eq.(8) allows only for an isoscalar central component $V_C$ and for isovector 
spin-spin and tensor components $W_{S,T}$.

The resulting NN-potentials in $r$-space, multiplied with $r \exp(3m_\pi r)$, 
are shown in Fig.\,6. The dominant feature is now a repulsive isoscalar 
central potential $\widetilde V_C(r)$, whereas the other potentials $\widetilde 
W_{S,T}(r)$ are suppressed by about one order of magnitude.     
 
\begin{figure}[h!]
\begin{center}
\includegraphics[scale=0.5,clip]{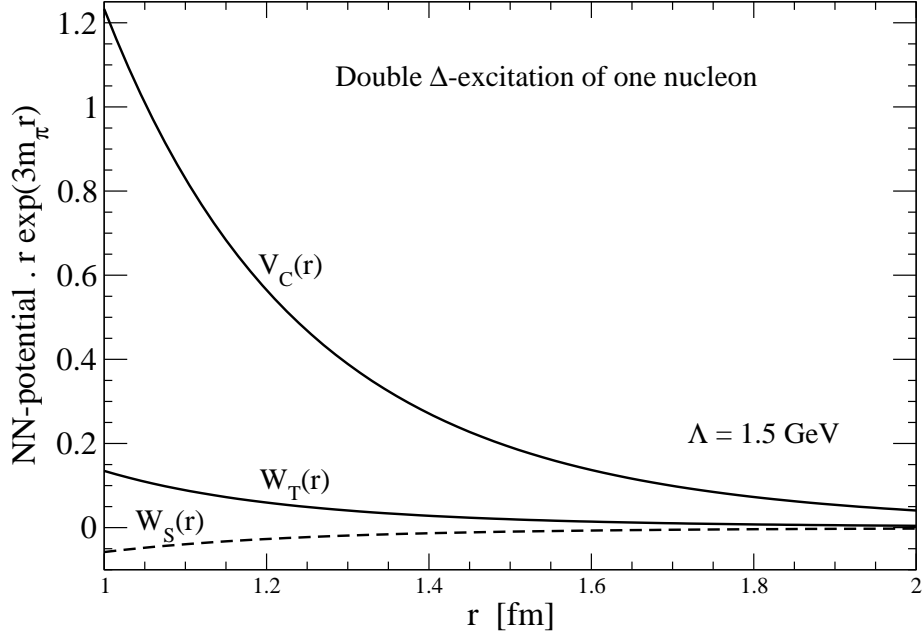}
\end{center}
\vspace{-.5cm}
\caption{NN-potentials arising from $3\pi$-exchange with double 
$\Delta$-excitation of one nucleon.}
\end{figure}

\begin{figure}[h!]
\begin{center}
\includegraphics[scale=0.5,clip]{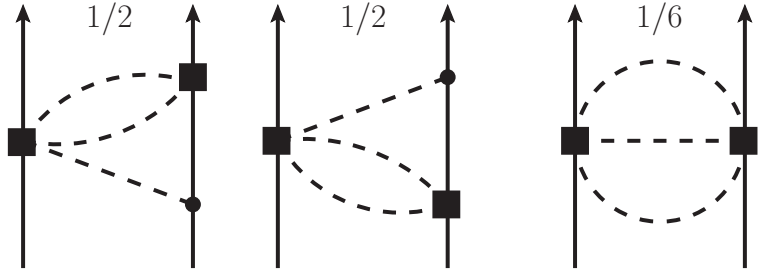}
\end{center}
\vspace{-.5cm}
\caption{The first two diagrams show $3\pi$-exchange with double and single 
$\Delta$-excitation of either one nucleon. The third diagram represents 
$3\pi$-exchange with double $\Delta$-excitation of both nucleons. The 
pertinent symmetry factors, $1/2$ and $1/6$, are indicated.}
\end{figure}
\begin{figure}[h!]
\begin{center}
\includegraphics[scale=0.5,clip]{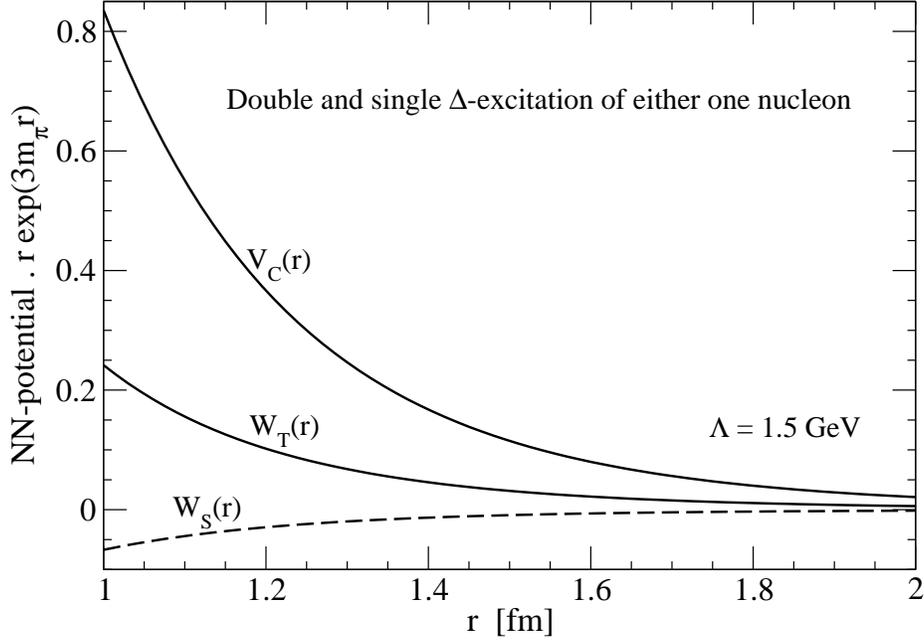}
\end{center}
\vspace{-.5cm}
\caption{NN-potential arising from $3\pi$-exchange with single and double 
$\Delta$-excitation of either one nucleon.}
\end{figure}

\subsection{Double and single $\Delta$-excitation of either one nucleon}
We continue with the $3\pi$-exchange mechanisms accompanied by double and 
single $\Delta(1232)$-excitation of either one nucleon. These are represented 
by the first two diagrams in Fig.\,7 (and their mirror partners). 
In this case the pertinent spectral-functions can be given in closed 
analytical form and they read:
\begin{equation} {\rm Im}V_C = {5g_A^6 m_\pi^7\over (4f_\pi)^6 \pi^2 \Delta^3} 
\bigg\{{R(u)\over 64}\bigg[{27\over u}+50-65u+12u^2+39u^3-2u^4-u^5\bigg] + 3L(u) \bigg[5u-{3u^3\over 2}-{1\over u}\bigg]\bigg\}\,, \end{equation}
\begin{eqnarray} {\rm Im}W_S &=&{g_A^6 m_\pi^7\over (4f_\pi)^6 \pi^2 \Delta^3} 
\bigg\{{R(u) \over 320}\bigg[{59u^5\over 10}+{59u^4\over 5}-{3377u^3\over 18}
+{4286u^2\over 45}+{20057u\over 30} \nonumber  \\ && +{307\over 3}-{37121\over 
90u}+{2488\over 45u^2}-{372\over 5u^3} \bigg]+L(u) \bigg[{u^3\over 4}-5u+
{40\over 3u}-{59\over 30u^3}\bigg] \bigg\}\,,\end{eqnarray}
\begin{eqnarray} {\rm Im}(\mu^2W_T-W_S) &=&{g_A^6m_\pi^7\over (4f_\pi)^6\pi^2 
\Delta^3} \bigg\{{R(u) \over 120}\bigg[{41u^5\over 20}+{41u^4\over 10}-{305u^3
\over 12}-{2141u^2\over 15}+{5543u\over 60}\nonumber  \\ &&
+{1621\over 6}+{29927\over 60u}+{622\over 15u^2}-{279\over 5u^3} \bigg]+
L(u) \bigg[2u^3-7u-{41\over 3u}-{59\over 15u^3}\bigg] \bigg\}\,,\end{eqnarray}
where the auxiliary functions $R(u)$ and $L(u)$ have been defined in eq.(9).

The resulting NN-potentials in $r$-space, multiplied with $r \exp(3m_\pi r)$, 
are shown in Fig.\,8. The dominant feature is again a repulsive isoscalar 
central potential $\widetilde V_C(r)$, while the other potentials $\widetilde 
W_{S,T}(r)$ are suppressed.
\subsection{Double $\Delta$-excitation of both nucleons}
Finally, there is the $3\pi$-exchange mechanism with double $\Delta(1232)
$-excitation of both nucleons. It is represented by the right diagram in 
Fig.\,7. While the imaginary part of the isoscalar central NN-potential  
Im$V_C$ can be given in terms of the short expression:  
\begin{equation} {\rm Im}V_C = -{25g_A^6 m_\pi^8\over (4f_\pi)^6 \pi^3 
(2\Delta)^4 u^2} \int\limits_2^{u-1}\!dw\,\big[(w^2-4)\lambda(w)\big]^{3/2}\,,
\end{equation}
rather lengthy formulas result from the multitude of spin-dependent terms in 
$M_{3\pi\Delta\Delta}$ written in eq.(8). The imaginary parts of the isovector 
spin-spin and tensor NN-potentials Im$W_{S,T}$ read:
\begin{eqnarray} {\rm Im}W_S &=& {g_A^6 m_\pi^8\over (5\pi)^3 f_\pi^6\Delta^4 
(4u)^8} \int\limits_2^{u-1}\!dw\,\sqrt{w^2-4}\,\lambda^{3/2}(w) \bigg\{
{\lambda^2(w)\over 42} +w^6\bigg({26u^2\over 9}+{1\over 21}\bigg)\nonumber\\ && 
+w^4\bigg({2065u^4\over 18}-{1487u^2\over 126}-{1\over 21}\bigg) +w^2\bigg(
{779u^6\over 18}-{20519u^4\over 42}+{349u^2\over 21}-{2\over 7}\bigg)\nonumber 
\\ &&  +{79u^8\over 18}-{243u^6\over 14}+{18545u^4\over 42}  
-{1633u^2\over 126}+{2\over 3}+{1\over 63 w^2}\big(556u^8-9358u^6 \nonumber\\ &&
+12199u^4+556u^2-33\big) +{1\over 21 w^4}\big(556u^8-1181u^6+697u^4-75u^2+3\big)
\bigg\}\,,\end{eqnarray}
\begin{eqnarray} {\rm Im}(\mu^2W_T-W_S) &=& {g_A^6 m_\pi^8\over (5\pi)^3f_\pi^6 
\Delta^4 (4u)^8} \int\limits_2^{u-1}\!dw\,\sqrt{(w^2-4)\lambda(w)}\, \bigg\{
{\lambda^3(w) \over 6} +{w^{10}\over 36}(403u^2+12)\nonumber\\ &&+{w^8\over 9}
\big(2494u^4-565u^2-9\big)-{w^6\over 18}\big(11707u^6+35060u^4-1993u^2+18\big)
\nonumber\\ && +{w^4\over 9}\big(3869u^8+29381u^6+42250u^4-423u^2+75\big)
\nonumber\\ && -w^2\bigg({2629u^{10} \over 36}+{10670u^8\over 9}+{32977u^6\over 
6}+{38701u^4\over 9}+{1997u^2\over 36}+15\bigg)\nonumber\\ && +13 +{u^2\over 9}
\big(47u^{10}-716u^8+9501u^6+29894u^4+ 8791u^2+430\big)\nonumber\\ &&  
+{1\over 9w^2} \big(97u^{12}+1920u^{10}-2013u^8 -2192u^6+2207u^4+32u^2-51\big)
\nonumber\\ && +{1\over 3w^4}\big(97u^{12}-400u^{10}+633u^8-472u^6+163u^4-24u^2+3
\big)\bigg\}\,.\end{eqnarray}

\begin{figure}
\begin{center}
\includegraphics[scale=0.5,clip]{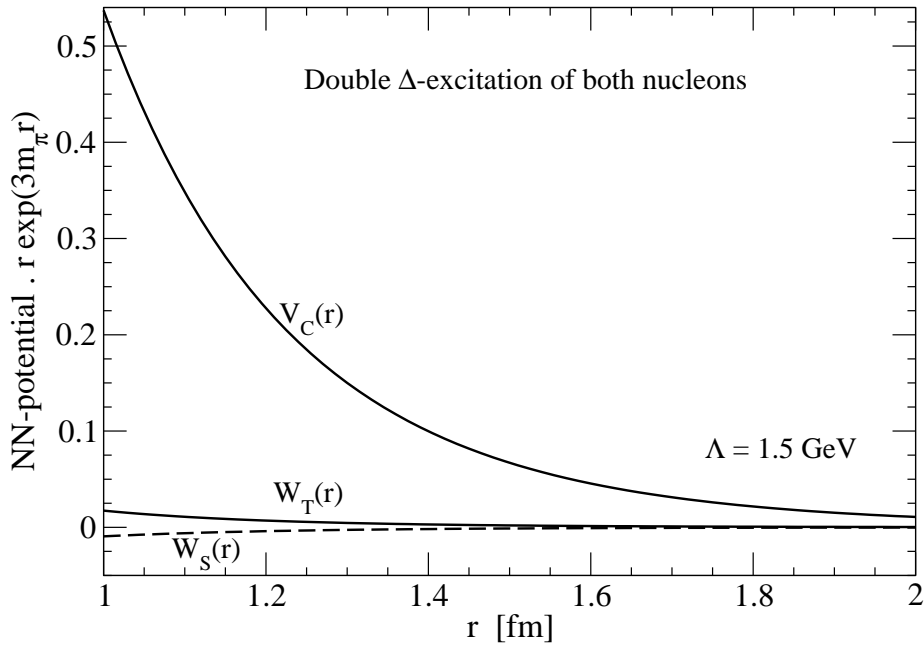}
\end{center}
\vspace{-.6cm}
\caption{NN-potentials arising from $3\pi$-exchange with double 
$\Delta$-excitation of both nucleons.}
\end{figure}
The corresponding NN-potentials in $r$-space, multiplied with $r 
\exp(3m_\pi r)$, are shown in Fig.\,9. The dominant feature is again a 
repulsive isoscalar central potential $\widetilde V_C(r)$. It is however 
smaller than those encountered in the previous two subsections.

\section{Conclusions}
\vspace{-0.2cm}
In this work we have studied the two-nucleon interaction which arises from 
the exchange of three pions and the excitation of $\Delta(1232)$-isobars in 
intermediate states. The pertinent two-loop spectral-functions have been 
calculated in the approximation of neglecting the energy-independence 
of the non-relativistic $\Delta(1232)$-propagator. A subsequent analysis 
of the NN-potentials in $r$-space has revealed that the main effect 
of these specific exchange and excitation mechanisms is isoscalar central 
repulsion. Other components, such as $\widetilde W_T(r)$ receive canceling 
contributions, or come out very small anyway. From the spectral-functions 
collected in section\,3 one can reconstruct the NN-potentials in momentum-space via 
subtracted dispersion relations. This way one can include the interaction terms 
calculated in this work (in  chiral effective field theory) into 
future fits to empirical NN phase shifts.      

\end{document}